\documentclass[useAMS,usenatbib]{mn2e}
\usepackage{epsfig}

\def\com{{\rm com}}
\def\lum{{\rm lum}}
\def\p{{\rm peak}}
\def\iso{{\rm iso}}
\def\obs{{\rm obs}}
\def\bol{{\rm bol}}

\title[Redshift Degeneracy in the $E_\iso$--$E_\p$ Relation] 
{Redshift Degeneracy in the $E_\iso$--$E_\p$ Relation of Gamma-Ray Bursts}
\author[Li-Xin Li]{Li-Xin Li\thanks{E-mail: lxl@mpa-garching.mpg.de}\\
Max-Planck-Institut f\"ur Astrophysik, 85741 Garching, Germany}

\begin{document}

\date{}

\date{Accepted 2006 October 13. Received 2006 October 11; in original form 2006
July 17}

\pagerange{\pageref{firstpage}--\pageref{lastpage}} \pubyear{2006}

\maketitle

\label{firstpage}

\begin{abstract}
In this {\it Letter} we show that there is a redshift degeneracy in the 
$E_\iso$--$E_\p$ relation of gamma-ray bursts (GRBs). If a GRB has a 
redshift solved from the $E_\iso$--$E_\p$ relation that lies in the range 
of $0.9<z<20$, a GRB that has the same observed fluence and peak spectral
energy but is at a different redshift in that range also satisfies the
$E_\iso$--$E_\p$ relation within 1-$\sigma$ error, implying an extremely
large error in the calculated redshift. Even if the data scatter in the 
$E_\iso$--$E_\p$ relation is reduced by a factor of 2, the error in the 
predicted redshift is still large enough to prevent from constraining the 
redshift meaningfully. Hence, the $E_\iso$--$E_\p$ relation is not useful 
for determining the GRB distance.
\end{abstract}


\section{Introduction}
\label{intro}

It has been found that the isotropic equivalent energy of long-duration 
gamma-ray bursts (GRBs) is correlated with the peak energy of their 
integrated spectra, with only a few outliers \citep{ama02,fri05,ama06}
\begin{eqnarray}
        E_\iso = A E_\p^m \;. \label{amati}
\end{eqnarray}
The peak energy $E_\p$ is defined to be the photon energy at the peak of 
the GRB spectrum $\nu F_\nu$, converted to the rest frame of the GRB. The 
isotropic equivalent energy $E_\iso$ is defined in the $1$--$10,000$ keV band 
in the GRB frame.

With a sample of 41 GRBs with firmly determined redshifts and peak spectral
energy, \citet{ama06} has found that $m\approx 2$ and $A\approx 0.92\times 
10^{48}$ erg ($E_\p$ in keV). The spread of the GRBs in the sample 
(1-$\sigma$ width) around the best fit model is given by $\sigma_{\log E_\p} 
\approx 0.15$, or equivalently, $\sigma_{\log E_\iso} \approx 0.3$. That
is, the uncertainty in the isotropic equivalent energy predicted by the 
relation in~(\ref{amati}) is about a factor of 2.

The $E_\iso$--$E_\p$ correlation~(\ref{amati}) is often called the Amati 
relation. A statistical analysis based on the correlation between the 
intensity and the peak spectral energy of GRBs with unknown redshifts 
carried out by \citet{llo00} before the discovery of the $E_\iso$--$E_\p$ 
relation \citep{ama02} has predicted the existence of such a relation and 
has indicated that it is a physical relation rather than a superficial 
relation.

The correlation is even better when the correction to the GRB energy from 
the jet collimation is included \citep{ghi04,ghi05,fri05}. However, this is 
achieved with a great price: the GRB jet opening angle is involved and hence 
the application of the relation requires multi-band observations on the 
afterglow as well as a knowledge on the jet model and the GRB environment 
parameters.

X-ray flashes (XRFs), a sub-class of GRBs that are characterized by a large
fluence in the X-ray energy band relative to the gamma-ray energy band 
\citep{sak04}, are found to satisfy the $E_\iso$--$E_\p$ relation. They 
include the extremely soft GRB 020903 \citep{sak04} and the recently 
discovered supernova-connected GRB 060218 \citep{cam06,ama06b}. 

Outliers to the $E_\iso$--$E_\p$ relation do exist. Short-duration GRBs are 
often found to violate the relation. Among long-duration bursts, famous 
outliers are GRB 980425 and GRB 031203 \citep{ama06}, both are sub-energetic 
and have spectroscopically confirmed supernovae accompanying them. 
\citet{nak05} and \citet{ban05} have argued that a significant fraction of 
the long-duration GRBs detected by BATSE on the {\it Compton Gamma-Ray 
Observatory} ({\it CGRO}) must violate the Amati relation. 

However, it has been claimed that GRB 031203 had a significant soft X-ray 
component not seen by the {\it International Gamma-Ray Astrophysics 
Laboratory} ({\it Integral}), which may make GRB 031203 consistent with the 
Amati relation by reducing the value of its $E_\p$ \citep{wat04,wat06,tie06}. 
This idea becomes particularly attractive after GRB 060218, which had a prompt 
emission directly detected in both gamma-ray and X-ray bands, lasting about 
2,000 seconds. It has motivated \citet{ghis06} to propose that GRB 980425 
and GRB 031203 are twins of GRB 060218 and they indeed satisfy the Amati 
relation.

Given its universality and simplicity, the Amati relation might be considered
as a good redshift or distance estimator for GRBs. In fact, it has already been
claimed by Amati (2006) that the relation can be the most reliable redshift 
estimator. Since it only involves the prompt emission of the GRB, one may hope
that the Amati relation can be applied to GRBs at very high redshift.

However, in this {\it Letter}, we show that it is impossible to use the 
$E_\iso$--$E_\p$ relation to estimate the redshift of a GRB at redshift $z\ga 
1$. This is because of the fact that there is an intrinsic redshift degeneracy
in the Amati relation, which causes more problems in determining the redshift
than the dispersion in the relation. The degeneracy arises from the definition
of the Amati relation and the fact that every empirical relation in astronomy 
has a nonzero intrinsic dispersion. The existence of the redshift degeneracy 
makes it impossible to distinguish GRB redshifts in the range of 
$z=0.9$--$20$, if 1-$\sigma$ error is allowed.

Throughout the {\it Letter}, we adopt a cosmology with $\Omega_m = 0.3$, 
$\Omega_\Lambda=0.7$, and $H_0 = 70$ km s$^{-1}$ Mpc$^{-1}$.

\section{The Redshift Degeneracy in the $E_\iso$--$E_\p$ Relation}
\label{degeneracy}

The peak spectral energy of the GRB, measured in the GRB frame, is related
to the  peak spectral energy determined in the observer's frame by
\begin{eqnarray}
	E_\p = (1+z) E_{\obs,\p} \;, \label{ep}
\end{eqnarray}
where $z$ is the redshift of the GRB.

The isotropic equivalent energy of the GRB, defined in the GRB frame, is 
related to the observed bolometric fluence of the GRB by 
\begin{eqnarray}
	E_\iso = \frac{4\pi D_\lum^2}{1+z}\; F_\bol
		= 4\pi D_\com^2 (1+z)\; F_\bol\;,
	\label{eiso}
\end{eqnarray}
where $F_\bol$ is the bolometric fluence, $D_\lum$ and $D_\com$ are, 
respectively, the luminosity distance and the comoving distance of the GRB. 

$D_\lum$ and $D_\com$ are calculated by
\begin{eqnarray}
	D_\lum = D_\com (1+z) = \frac{c}{H_0} (1+z)\, w(0,z) \;, 
	\label{dlum} 
\end{eqnarray}
where $c$ is the speed of light, $H_0 = 100 h$ km s$^{-1}$ Mpc$^{-1}$ is 
today's Hubble constant, and
\begin{eqnarray}
	w(z_1,z_2) \equiv \int_{z_1}^{z_2} \frac{dz}{\sqrt{\Omega_m
		(1+z)^3 + \Omega_R(1+z)^2 + \Omega_\Lambda}} \;.
	\label{w}
\end{eqnarray}
Here, $\Omega_m$, $\Omega_\Lambda$, and $\Omega_R = 1-\Omega_m - 
\Omega_\Lambda$ are, respectively, the ratio of today's mass density in dark 
and ordinary matter, in vacuum energy (cosmological constant or dark energy), 
and in the spatial curvature of the universe to the critical mass density 
\citep{pee03}.

\begin{figure}
\vspace{2pt}
\includegraphics[angle=0,scale=0.602]{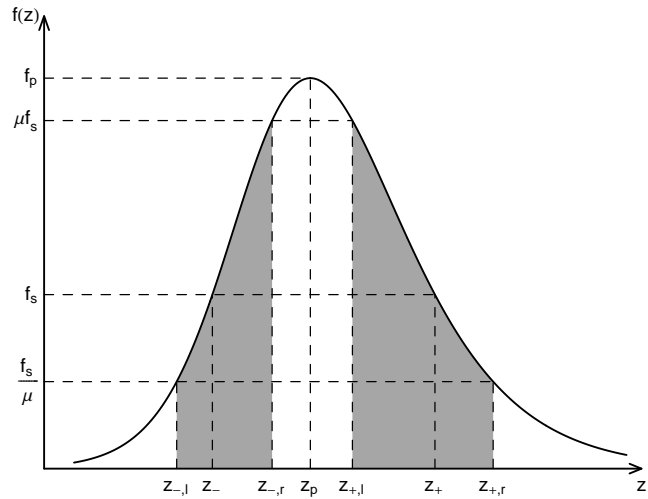}
\caption{The function $f(z)$ and the solution for the GRB redshift ($z_\pm$)
determined by eq.~(\ref{eq_sol}). $f_s = f(z_-) = f(z_+)$, which is equal 
to the right-hand side of eq.~(\ref{eq_sol}). The solutions of $f(z) =
\mu f_s$ and $f(z) = f_s/\mu$ determine the 1-$\sigma$ regions of $z_\pm$
(eqs.~\ref{sigma1}, \ref{sigma2}; shaded regions). $f_p$ is the maximum of 
$f(z)$, occurring at $z=z_p$. When $f_s > f_p$, eq.~(\ref{eq_sol}) has no 
solution. (The redshift $z$ is in logarithmic scale.)
}
\label{z_sol}
\end{figure}

If a GRB has an observed bolometric fluence $F_\bol$ and an observed peak 
spectral energy $E_{\obs,\p}$ but an unknown redshift, we can solve for the 
redshift from the $E_\iso$--$E_\p$ relation (eq.~\ref{amati}), by assuming 
that the GRB satisfies the relation. The dispersion of the $E_\iso$--$E_\p$ 
relation will lead to an error in the calculated redshift.

Submitting equations~(\ref{ep}) and (\ref{eiso}) into equation~(\ref{amati}),
we get
\begin{eqnarray}
	f(z) \equiv \frac{w(0,z)^2}{(1+z)^{m-1}} 
		= \frac{A}{4\pi} \left(\frac{c}{H_0}\right)^{-2}
		\frac{E_{\obs,\p}^m}{F_\bol} \;.
	\label{eq_sol}
\end{eqnarray}
All the redshift dependence is on the left-hand side.

Equation~(\ref{eq_sol}) does not always have a solution. If $m>1$ ($m\approx 
2$ in the Amati relation), the function $f(z)$ approaches zero as 
$z\rightarrow 0$ and $z\rightarrow \infty$, and peaks at a redshift $z_p$. 
The maximum of $f$ is then $f_p = f(z_p)$. If the right-hand side of
equation~(\ref{eq_sol}), which we denote by $f_s$, is greater than $f_p$, 
solution for $z$ will not be found. Based on this kind of arguments, 
\citet{nak05} have shown that at least 25\% of the bursts in the BATSE 
sample are outliers to the Amati relation (see also Band \& Preece 2005).

On the other hand, if $f_s<f_p$, we will find two solutions for the 
redshift, which we denote by $z_-$ and $z_+$, $z_-<z_+$. In the limiting 
case of $f_s=f_p$, we have $z_- = z_+$.

\begin{figure}
\vspace{2pt}
\includegraphics[angle=0,scale=0.755]{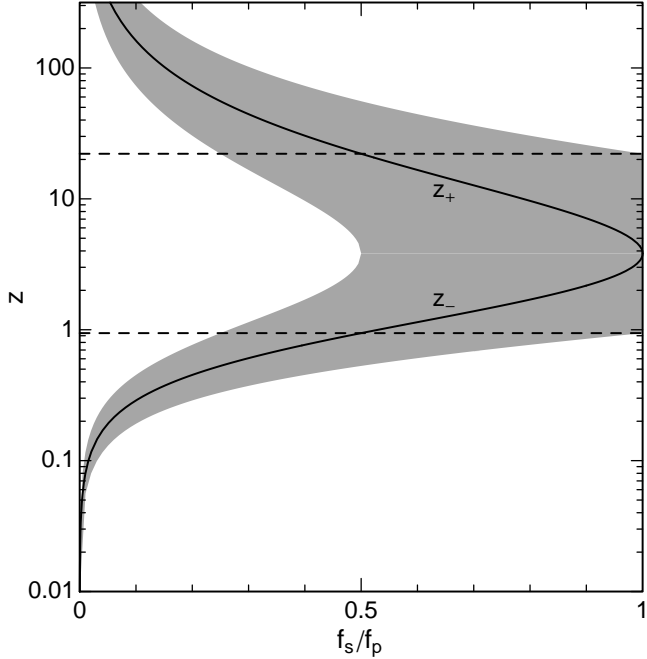}
\caption{The redshift determined by the Amati relation (i.e., the solutions
to eq.~\ref{eq_sol}, with $m=2$), and 1-$\sigma$ error. $f_s$ is equal to 
the right-hand side of eq.~(\ref{eq_sol}), and $f_p$ is the peak of $f(z)$ 
(Fig.~\ref{z_sol}). When $f_s>f_p$, eq.~(\ref{eq_sol}) has no solution. When 
$f_s<f_p$, eq.~(\ref{eq_sol}) has two solutions: $z_-$, the lower branch of 
the solid curve; and $z_+$, the upper branch. The two solutions merge at 
$f_s=f_p$. The shaded region shows the 1-$\sigma$ error of $z_\pm$, arising 
from the spread in the Amati relation (a factor of $\mu=2$ in the isotropic 
energy). When $f_s\ge f_p/\mu = 0.5 f_p$, the 1-$\sigma$ region of $z_-$ 
merges with that of $z_+$, leading to a single 1-$\sigma$ region in $z_\pm$ 
that includes the entire region of $z$ bounded by the two horizontal dashed
lines (defined by $z=0.94$ and $z=22$). 
}
\label{z_fs}
\end{figure}

The error in $z_\pm$ is determined by the spread in the Amati relation.
Assume that the spread is given by a factor of $\mu$ ($>1$) in the calculated 
isotropic energy, i.e., within 1-$\sigma$ the isotropic energy of the GRB 
is expected to be in the range of $E_\iso/\mu$--$\mu E_\iso$, where 
$E_\iso$ is the isotropic energy calculated with equation~(\ref{amati}).
Based on a sample of 41 long-duration GRBs, \citet{ama06} obtained that $\mu 
\approx 2$ (see the 2nd paragraph in Sec.~\ref{intro} of this {\it Letter}).

\begin{figure}
\vspace{2pt}
\includegraphics[angle=0,scale=0.73]{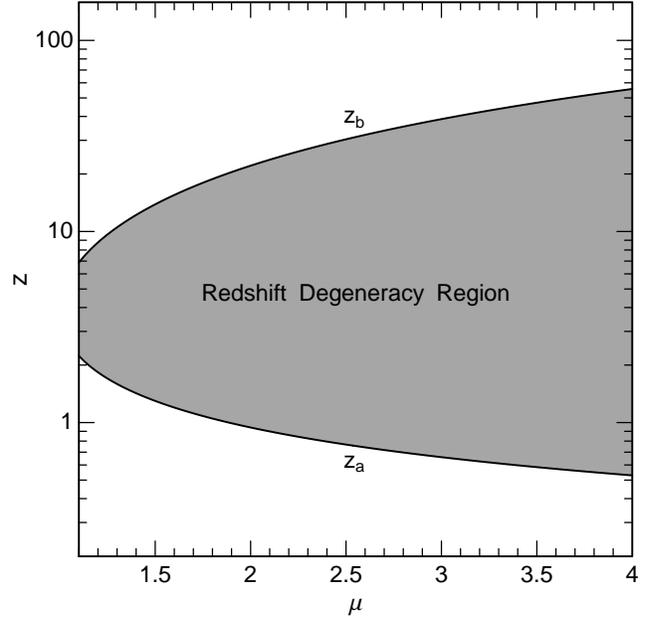}
\caption{The 1-$\sigma$ redshift degeneracy region (the shaded region) 
bounded by $z_a$ and $z_b$, obtained by solving eq.~(\ref{zab}). Given
the 1-$\sigma$ spread in the $E_\iso$--$E_\p$ relation---a factor of 
$\mu = 2$ in $E_\iso$ obtained with the current data---the redshift 
degeneracy region spans from $z=0.94$ to $z=22$.
}
\label{z1_z2}
\end{figure}

Imagine that, the same observed GRB that satisfies equation~(\ref{amati}) 
at $z=z_\pm$, is assigned another redshift $z^\prime$. Then, the GRB would
have an isotropic equivalent energy
\begin{eqnarray}
	E_\iso^\prime = E_\iso \frac{D_\lum^{\prime 2}}{D_\lum^2}
		\frac{1+z}{1+z^\prime} \;, \nonumber
\end{eqnarray}
and a peak spectral energy
\begin{eqnarray}
	E_\p^\prime = E_\p \frac{1+z^\prime}{1+z} \;. \nonumber
\end{eqnarray}
Thus we have
\begin{eqnarray}
	E_\iso^\prime = A E_\p^{\prime m} \frac{f(z^\prime)}{f(z)} \;.
\end{eqnarray}
The 1-$\sigma$ ranges of $z_-$ and $z_+$, denoted respectively by 
$(z_{-,l}, z_{-,r})$ and $(z_{+,l}, z_{+,r})$, are hence determined by
\begin{eqnarray}
	f(z_{-,l}) = f(z_-)/\mu \;, \hspace{1cm}
	f(z_{-,r}) = \mu f(z_-) \;; \label{sigma1}
\end{eqnarray}
and
\begin{eqnarray}
	f(z_{+,l}) = \mu f(z_+) \;, \hspace{1cm}
	f(z_{+,r}) = f(z_+)/\mu \;. \label{sigma2}
\end{eqnarray}

The solutions of $z_\pm$, as well as their 1-$\sigma$ ranges, are sketched
in Fig.~\ref{z_sol}. Obviously, the 1-$\sigma$ regions of $z_-$ and $z_+$ 
overlap if $f_s = f(z_\pm) \ge f_p/\mu$.

For $m=2$ (then $z_p=3.825$, $f_p=0.561$, for $\Omega_m = 0.3$ and 
$\Omega_\Lambda = 0.7$), $\mu = 2$ \citep{ama06}, and $f_s \le f_p$, the 
solutions of $z_\pm$ and their errors are shown in Fig.~\ref{z_fs}. A 
striking feature is that for $f_p/\mu\le f_s\le f_p$, the 1-$\sigma$ regions 
of $z_\pm$ merge to form a single 1-$\sigma$ region that spans a very 
large interval in redshift, covering the entire range from $z=0.94$ to 
$z=22$ (the region bounded by the two dashed lines). This means that, if the 
redshift of a GRB calculated by the Amati relation is in the range of 
$0.94$--$22$, then within 1-$\sigma$ error the redshift can have any value 
in the range from $0.94$ to $22$. Thus, $0.94<z<22$ is a {\it redshift 
degeneracy} region since in which the redshift of a GRB cannot be determined 
by the Amati relation if 1-$\sigma$ error is allowed. If 2-$\sigma$ error 
($\mu = 2^2 =4$) is allowed, then the  redshift degeneracy region is 
$0.53<z<55$. Even if the scatter in the Amati relation is reduced by a 
factor of 2, i.e., $\mu = 2^{1/2} = 1.4$, the redshift-degeneracy region 
is $1.4<z<12$ in 1-$\sigma$, still quite large. 

\begin{figure}
\vspace{2pt}
\includegraphics[angle=0,scale=0.73]{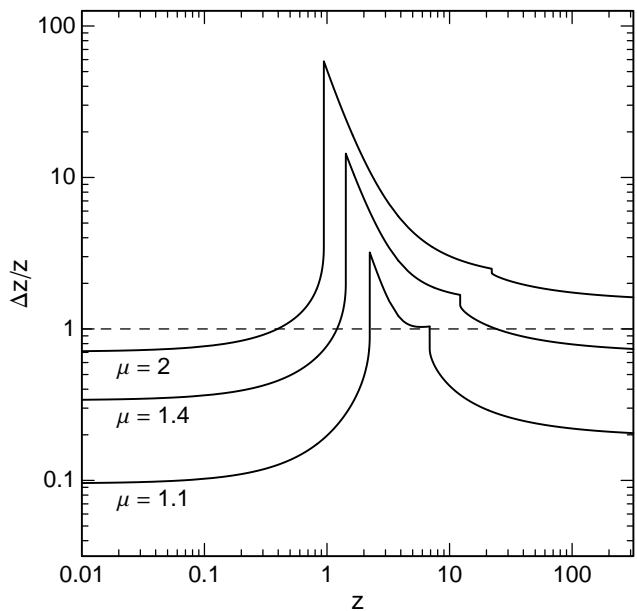}
\caption{The fractional error in the GRB redshift determined by the Amati
relation, as a function of the solved redshift. Different curves correspond to
different choice of $\mu$, as labeled. The horizontal dashed line marks
the position of $\Delta z = z$.
}
\label{ze_z}
\end{figure}

The merged 1-$\sigma$ region is bounded by $z_{-,l}$ and $z_{+,r}$, and is 
minimized when $f_s = f_p$. Thus, with 1-$\sigma$ error allowed, the redshift 
degeneracy region is bounded by the two roots of the equation
\begin{eqnarray}
	f(z) = f_p/\mu \;. \label{zab}
\end{eqnarray}
The two roots, $z_a$ and $z_b$ as functions of $\mu$, as well as the redshift
degeneracy region bounded by them, are shown in Fig.~\ref{z1_z2}.

The width of the 1-$\sigma$ region of the solution $z=z_\pm$ is $\Delta z
= z_r - z_l$, where $z_l$ and $z_r$ define the boundary of the 1-$\sigma$ 
region. If $f_s<f_p/\mu$, the 1-$\sigma$ region of $z_-$ and that of $Z_+$ are
separated, so that $z_l = z_{-,l}$, $z_r = z_{-,r}$ for $z=z_-$, and $z_l = 
z_{+,l}$, $z_r = z_{+,r}$ for $z=z_+$. If $f_p/\mu \le f_s \le f_p$, the two 
1-$\sigma$ regions merge to form a single one with $z_l = z_{-,l}$ and $z_r = 
z_{+,r}$ for both $z=z_-$ and $z=z_+$, which causes a jump in $\Delta z$ at 
$f_s = f_p/\mu$.

If $\Delta z >z$, the redshift solved from the Amati relation will not be 
very useful since the error in it is too large. In Fig.~\ref{ze_z} we plot 
the ratio $\Delta z/z$ against $z$, for different choice of the spread 
parameter $\mu$. For $\mu=2$, inferred from the current data \citep{ama06}, 
$\Delta z/z > 1$ when $z>0.395$ and up to at least $z\approx 300$ the upper 
limit plotted in the figure. When $\mu=1.4$ (i.e., the spread in $\log 
E_\iso$--$\log E_\p$ is reduced by a factor of 2 compared to $\mu = 2$), 
$\Delta z/z > 1$ when $1.21<z<24$. When $\mu=1.1$ (i.e., the spread in 
$\log E_\iso$--$\log E_\p$ is reduced by a factor of 7.3 compared to $\mu 
= 2$!), $\Delta z/z > 1$ when $2.2<z<6.9$.

\section{Conclusions}
\label{concl}

We have shown that the $E_\iso$--$E_\p$ relation has a serious redshift
degeneracy which prohibits from using the $E_\iso$--$E_\p$ relation with
the spread given by the current data to determine the redshift of a GRB in 
the range of $z=0.9$--$20$. Even if the spread in the relation is reduced 
by a factor of 2, the redshift degeneracy region is still quite larger, 
$z=1.4$--$12$. The numbers cited above are obtained under the assumption 
that 1-$\sigma$ error is allowed. If 2-$\sigma$ error is allowed, the 
redshift degeneracy region in the current $E_\iso$--$E_\p$ relation is 
$z=0.5$--$50$.

Mathematically, the existence of the redshift degeneracy is caused by the 
fact that in the $\log E_\iso$--$\log E_\p$ plane, the trajectory of a GRB is 
almost parallel to the $E_\iso$--$E_\p$ relation, if the redshift of the GRB 
varies from $z\sim 1$ to $z\sim 20$ but the observed fluence and peak energy
remain unchanged \citep{nak05,don06}. Thus, if a GRB satisfies the Amati 
relation, as its redshift varies in this range the GRB remains in the 
neighborhood of the relation within 1-$\sigma$ error. This is most clearly 
described by the case of GRB 060121, whose redshift is constrained to lie 
between $z=1.5$ and $4.6$ by its optical and near infrared afterglow 
observations \citep{don06}. As the redshift of GRB 060121 varies from $z=1.5$ 
to $z=5$, the burst always satisfies the Amati relation (Donaghy et al. 2006, 
Fig.~19).

Although the degeneracy described in this {\it Letter} does not exist for
GRBs at low redshift, the error in the low redshift determined by the Amati 
relation is still large. For example, if the GRB is at $z=0.1$, the redshift
solved from the Amati relation is $0.1_{-0.03}^{+0.05}$ ($\mu =2$). Because 
of this and the additional fact that the corresponding relation for Type Ia 
supernovae has a much smaller scatter, GRBs are not very useful for 
cosmology at low redshift.

A big problem that has hindered the development in understanding the nature
of GRBs has been the determination of the distance of GRBs. Various redshift 
and distance estimators for GRBs have been proposed, based on the 
empirical correlations that have been discovered \citep{mez06}. The results 
in this {\it Letter} indicate that the $E_\iso$--$E_\p$ relation cannot be 
used to determine the redshift and distance of GRBs.

\section*{Acknowledgments}

The author thanks Bohdan Paczy\'nski for stimulating discussions and helpful
comments. He also thanks the referee Darach Watson for a helpful report.

\bsp

\label{lastpage}

\end{document}